\documentclass{jaa}
\usepackage{natbib}
\usepackage{aas_macros}

\usepackage{bm}

\usepackage{graphicx}
\usepackage{siunitx}
\usepackage[colorlinks=true,linkcolor=black,anchorcolor=black,citecolor=black,filecolor=black,menucolor=black,runcolor=black,urlcolor=black]{hyperref} % Adds clickable links at references
\usepackage{amsmath}
\usepackage[bottom]{footmisc}
\usepackage{chngcntr}

\usepackage{dirtytalk} % for quotations
\usepackage{multicol} % for rcw results
\usepackage{subcaption} % for subfigures
\usepackage{makecell} % for our table style

\usepackage[utf8]{inputenc}
\DeclareUnicodeCharacter{03BC}{\ensuremath{\mu}}

\usepackage{graphicx}%
\usepackage{multirow}%
\usepackage{amsmath,amssymb,amsfonts}%
\DeclareMathSymbol{\mathsemicolon}{\mathpunct}{operators}{"3B} % Define semicolon as a math symbol
\usepackage{comment}
\usepackage{xcolor} 

\usepackage{hyperref}
\hypersetup{
    colorlinks=true,
    linkcolor=blue,
    filecolor=magenta,      
    urlcolor=cyan,
    citecolor=blue,
    pdftitle={VolDen_JOAA_v1},
    pdfpagemode=FullScreen,
    }

\usepackage{mathrsfs}%
\usepackage[title]{appendix}%
\usepackage{xcolor}%
\usepackage{textcomp}%
\usepackage{manyfoot}%
\usepackage{booktabs}%
\usepackage{algorithm}%
\usepackage{algorithmicx}%
\usepackage{algpseudocode}%
\usepackage{listings}%
\usepackage{gensymb}

\newcommand\arcsec{\mbox{$^{\prime\prime}$}}%
\newcommand\farcs{\mbox{$.\!\!^{\prime\prime}$}}%

\newcommand{\volden}{{\fontfamily{qcr}\selectfont VolDen}}
\newcommand{\radfil}{{\fontfamily{qcr}\selectfont RadFil}}
\newcommand{\FILFINDER}{{\fontfamily{qcr}\selectfont FILFINDER}}
\newcommand{\DISPERSE}{{\fontfamily{qcr}\selectfont DISPERSE}}
\newcommand{\FELLWALKER}{{\fontfamily{qcr}\selectfont FELLWALKER}}
\newcommand{\RNum}[1]{\uppercase\expandafter{\romannumeral #1\relax}}

\begin{document}\sloppy

%%paper title
%%For line breaks \\ can be used within title
\title{\volden: a tool to extract number density from the column density of filamentary molecular clouds}

%%author names are separated by comma (,)
%%use \and before the last author name
%%use a * along with the number separated by comma
%% for the  author for correspondence
%%\textsuperscript{number} is used for affiliation
%%\affilOne, \affilTwo etc., upto \affilTwentyfive is possible
%%Please note the first letter after \affil is capitalised in the command
%%

\author{Ashesh A. K\textsuperscript{1}, Chakali Eswaraiah\textsuperscript{1,*} and P Ujwal Reddy\textsuperscript{1}, Jia-Wei Wang\textsuperscript{2}}
\affilOne{\textsuperscript{1}Department of Physics, Indian Institute of Science Education and Research (IISER), Tirupati 517507, India.\\}
\affilTwo{\textsuperscript{2}Academia Sinica Institute of Astronomy and Astrophysics, Taipei 10617, Taiwan.}

%%escape two column mode for title, affiliation and abstract
%%by giving \twocolumn command as shown

\twocolumn[{
%\onecolumn{

\maketitle

%%include \corres to print the corresponding author Email id
\corres{eswaraiahc@labs.iisertirupati.ac.in}

%%include \msinfo for
%%manuscript information such as
%%received, revised and accepted dates
%%

%\msinfo{1 January 2015}{1 January 2015}

%%abstract
\begin{abstract}
Gas volume density is one of the critical parameters, along with dispersions in magnetic field position angles and non-thermal gas motions, for estimating the magnetic field strength using the Davis-Chandrasekhar-Fermi (DCF) relation or through its modified versions for a given region of interest. We present \volden~an novel python-based algorithm to extract the number density map from the column density map for an elongated interstellar filament. \volden~uses the workflow of \radfil~to prepare the radial profiles across the spine. The user has to input the column density map and pre-computed spine along with the essential \radfil~parameters (such as distance to the filament, the distance between two consecutive radial profile cuts, etc.) to extract the radial column density profiles. The thickness and volume density values are then calculated by modeling the column density profiles with a Plummer-like profile and introducing a cloud boundary condition. The cloud boundary condition was verified through an accompanying N-PDF column density analysis. In this paper, we discuss the workflow of~\volden~and apply it to two filamentary clouds. We chose LDN\,1495 as our primary target owing to its nearby distance and elongated morphology. In addition, the distant filament RCW\,57A is chosen as the secondary target to compare our results with the published results. Upon publication, a complete tutorial of \volden and the codes will be available via \href{https://github.com/aa16oaslak/volden}{GitHub}.
\end{abstract}

%%insert keywords separated by 3 hyphens using \keywords{words}
\keywords{ISM: clouds, filaments, magnetic fields -- Astronomy: data analysis}

%}
%%close the one column escape here

}]
%%close the twocolumn escape here

%%include \doinum{number}for the DOI number in the header
%%include \volnum{number} for the volume number in the header
%%include \year{yyyy} for  year of publication in the header
%%include \pgrange{num--num} page range of article in the header
%%include \artcitid{num} for the article citation id
%%include \lp to print last page of the article
%%include \setcounter{page}{pagenum} for the exact starting page of the article

\doinum{12.3456/s78910-011-012-3}
\artcitid{\#\#\#\#}
\volnum{000}
\year{0000}
\pgrange{1--}
\setcounter{page}{1}
\lp{1}

\section{Introduction}\label{intro}

Filamentary structures are found to be ubiquitous in the molecular clouds as per the dust emission observations in the far-infrared and sub-millimeter wavelengths by \textit{Herschel} space observatory  \citep{Andr__2010, Menshchikov_etal_(2010), Arzoumanian_2011, hill_etal.(2011)} as well as by gas tracers \citep{schneider1979catalog, johnstone1999scuba,myers2009distribution}. These filaments fragment into dense cores, which appear in a chain-like fashion along the dense ridges, and they further collapse to form protostars \citep{Andr__2010, Arzoumanian_2011, konyves2015census,marsh2016evidence}. Both observational and simulation results suggest a direct link between the filamentary structures and the star formation processes \citep{Andr__2010,Federrath_2016}. 
The origin of these filamentary structures is generally attributed to the complex interplay between gravity, turbulence, and magnetic fields (B-field) \citep{mckeeostriker2010, Soler_2013, Li_Klein_2019}.

One of the most commonly used tools for tracing the orientation of plane-of-the-sky (POS) B-fields ($B_{\mathrm{pos}}$) is by observing the polarized light from the background stars in optical and near-infrared wavelengths \citep{Pereyra_2004, sugitani_etal_2011, franco&alves_2015, Eswaraiah_2017} and polarized thermal dust emission from the dense cloud/core regions in far-infrared and sub-millimeter wavelengths,  \citep{girart2006, Matthews_2009, dotson2010, Hull_2013, Planck_Collaboration_2015, Cortes_etal_2016, Chuss_etal_2019}. The origin of observed polarization has been attributed to the aligned interstellar dust grains with respect to B-fields \citep{hall1949,hiltner1949b,greenstein1951}). Subsequently, the POS component of magnetic field strength is proposed to be estimated by an indirect but statistical approach, famously known as the Davis-Chandrasekhar-Fermi (DCF) method \citep{Chandrasekhar_fermi}:

\begin{equation}
B_{pos} = Q \sqrt{4\pi\rho} \frac{\delta_{V}}{\delta_{\theta_{B}}},
\end{equation}
where Q is the constant factor 0.5 \citep{Ostriker_2001}, $\delta_{\theta_{B}}$ is the {dispersion in the polarization angles}, $\delta_{V}$ is the non-thermal or turbulent gas velocity dispersion and mass density $\rho = \mu m_{H}n(H_{2})$; $\mu$ is the mean molecular weight, $m_{H}$ is the mass of the hydrogen atom, and $n(H_{2})$ is the gas number density. The DCF method is based on the assumption 
 that isotropic turbulent motions are incompressible and manifest as propagating Alfv\'{e}nic waves. However, a recent work treated turbulence as anisotropic and compressible; hence, non-Alfv\'{e}nic modes may be prominent \citep{Skalidis_Raphael_2021}. They have suggested the relation: $B_{pos} = \sqrt{2\pi\rho} \frac{\delta_{V}}{\sqrt{\delta_{\theta_{B}}}}$ (hereafter ST method). In this paper, we have used both DCF and ST methods to calculate the POS B-field strengths. 

The gas number density is one of the prime factors, other than dispersions in both magnetic fields and gas non-thermal motions employed in DCF and ST methods.  Dispersion in magnetic fields can be estimated using simple Gaussian mean over a distribution of observed ordered polarization angles \citep[e.g.,][]{Santos_et_al_2014, Eswaraiah_2019}. Subsequently, more sophisticated means such as structure-function \citep{Hildebrand_2009} and autocorrelation function analyses \citep{Houde_2009} are adopted, especially to directly estimate the turbulent to ordered magnetic field strengths. To account for the non-thermal or turbulent component of the molecular gas, the gas velocity dispersion is extracted from the spectral data of molecular lines of a reliable gas tracer for a given density and by adequately accounting for the thermal contribution. The column density map can be used to extract the gas number density; however, due to the uncertain geometry of the cloud, this is not always straightforward. Filamentary clouds have varying geometry, implying that the depth of the cloud varies with gas density. As a result, a single approach cannot be used to extract the number density from column density at different scales and densities of molecular clouds.

A small fraction of the material in a cloud lies in the form of dense cores ($<$ 10\%; \citealt{Enoch_2007}), which are shown to be the primary sites of the star formation process. 
Most approaches for calculating gas volume density focus on the dense cores of the filamentary clouds. For instance, \citet{Liu_2018} set a threshold of $N_{\mathrm{H_{2}}} > 7 \times 10^{21} \mathrm{cm^{-2}}$ (defined by \citealt{Andr__2010} for nearby clouds) for identifying the dense cloud clumps using the \FELLWALKER\footnote{\href{http://www.starlink.ac.uk/docs/sun255.htx/sun255se2.html\#x3-60001}{Link to \FELLWALKER}} algorithm. The dense cores are assumed to be of spherical, oblate, or prolate geometry and are modeled as ellipses based on their two-dimensional distribution of dust emission on POS. Then, a geometric radius of the core was estimated using the relation $R_{\mathrm{c}} = \sqrt{ab}$, where $a$ and $b$ are the extents of major and minor axes of the cores. The ratio of the total mass ($M_{\mathrm{c}}$) to its volume ($V_{\mathrm{c}}$) gives the mass density and hence the number density of the cores

\begin{equation}
    \begin{aligned}[b]
    \rho &= \frac{M_{c}}{V_{c}},\,\,\, \rho = \mu m_{H} n(H_{2}), \\
    %\implies
    \mathrm{and} \,\,\,  
    n(H_{2}) &= \frac{M_{c}}{V_{c}\mu m_{H}} = \frac{M_{c}}{\frac{4}{3}\pi R_{c}^3\mu m_{H}}.
    \end{aligned}
\label{liu2018_voldensity_eqn}
\end{equation}

Similarly, other studies \citep[e.g.,][]{Anindya_saha_2022, jiao_etal, fengwei_etal2023(ATOMS)} have followed the same approach by assuming that the dense cores are \say{\textit{spherical}} and dust emission from them is optically thin and follows the modified Planck function at a given mean dust temperature \citep{hildebrand1983determination}.

Since the geometry and density of different portions of a cloud (such as filaments, clumps, and cores) are not the same, a single method of obtaining number density by simply dividing column density by the spatial extent of the structure can not be applied. Therefore, it is crucial to determine the depth and geometry of both low-density extended regions and dense filaments (and their fragments). To obtain the number densities for both low- and high-density regions (or large- to small-scales) of a cloud at once, we have created \volden, a python-based pipeline to extract the gas volume density of filaments having curved morphology like Taurus LDN\,1495 filament (L1495). This study assumes that the filament lies on the plane-of-the-sky (POS) and follows a symmetric cylindrical morphology across the spine. We used \radfil~\citep{Zucker_2018} to build radial profiles to preserve the filament's morphology across the spine. We have used the Plummer-like model to fit the extracted column density profiles. Since the Plummer-like model represents a filamentary cylindrical structure having no fixed boundary, we used a cloud boundary condition as defined by \citet{Wang_2019} to construct a thickness map of the filament (discussed in section \ref{Proposed technique}). Subsequently, we employed an {column density probability distribution function (N-PDF) analysis} to check our cloud boundary assumption. Finally, the number density map of the filament has been created using the column density and thickness map. Section \ref{contributing_data} discusses the data we have used for our analysis. In section \ref{methods}, we reviewed the workflow of \volden. Section \ref{taurus_results} presents the results based on the application of~\volden~for L1495. In section \ref{concept_proof}, as a proof of concept, we summarize the results on magnetic field strength for RCW\,57A obtained using \volden ~and compared it with the published studies \citep{Eswaraiah_2017}.

\section{Data}\label{contributing_data}

\begin{figure}%[!ht]
\centering
\includegraphics[width=9.25cm, height =9.25cm]{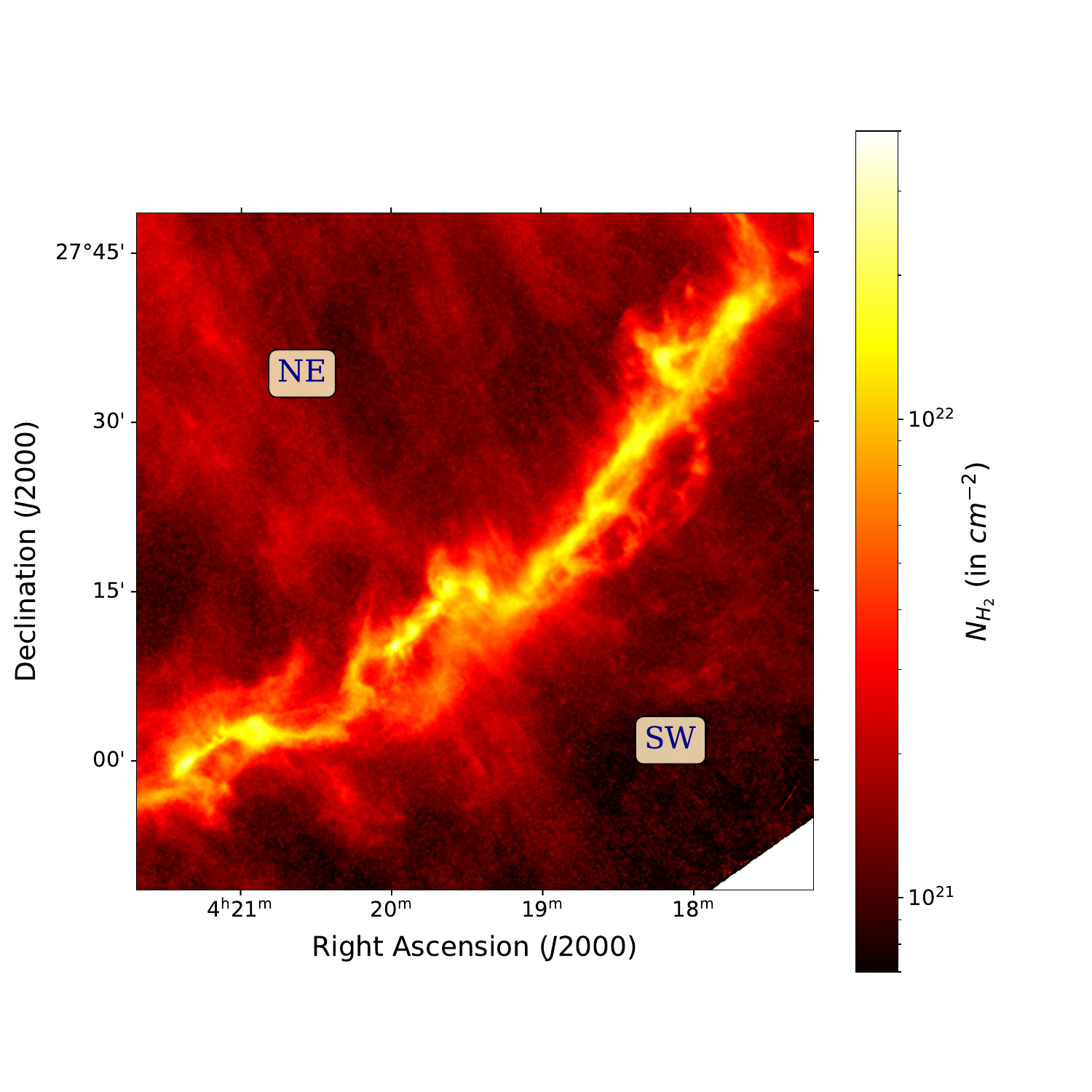}%[width=0.7\textwidth]{taurus_plots/taurus_column_density_map.pdf}%0.6\textwidth]{taurus.eps}
\caption{The  high-resolution column density map of the Taurus region obtained from the \textit{Herschel} data \citep{Palmeirim_2013}.}
\label{fig1}
\end{figure}

{The Gould Belt model represents an expanding ring structure composed of young massive stars classified as O- or early-type B (OB stars), associated molecular clouds, and neutral hydrogen located within a distance of 500 pc from the Sun, and tilted at an angle of approximately 20 degrees relative to the Galactic plane \citep{Palous2017}. Over the past 150 years, the prevailing framework for understanding the local interstellar medium has been based on this model. This model is not perfect, and research is still ongoing to understand the anomalies associated with this model-- e.g., \citet{alves_2020Nature} analyzed the 3-D structure of the local cloud complex and have shown that many of the local dense cloud structures contain multiple clouds that were initially thought to be part of the Gould Belt. In this study, we have used the high-resolution column density map obtained from Herschel Gould Belt Survey\footnote{\href{http://www.herschel.fr/cea/gouldbelt/en/Phocea/Vie_des_labos/Ast/ast_visu.php?id_ast=66}{\textit{Herschel} data archive}} \citep{Andr__2010}.} Herschel performed extensive far-infrared and sub-millimeter mapping of nearby molecular clouds that are part of the Gould Belt with SPIRE at 250~--~500 $\mu$m and PACS at 110~--~170 $\mu$m. The 15 arcsec angular resolution ($\lambda \approx 200 \mu m$) of {\it Herschel} is suitable for observing individual star-forming cores up to a distance of 0.5 kpc, thus making Herschel the best survey to characterize the early stages of star formation processes in the cloud complexes of Gould Belt \citep{Andr__2010}. The high-resolution column density map ($18\farcs2$) of the Taurus region produced by \citet{Palmeirim_2013} was used to demonstrate the performance of \volden. Since Taurus is located at 140 pc \citep{elias_1978, Galli_2018}, the column density map provides a detailed view of L1495 (see Figure~\ref{fig1}). Here, we mainly focus on the L1495 filament and nearby area, which is a part of the 10-pc-long large-scale filament L1495 in the Taurus molecular cloud  \citep{Palmeirim_2013,tafalla_etal_2015}.  It is an elongated filament with a curved morphology (Figure~\ref{fig1}), making it an ideal target for testing our algorithm. In this work, we publish the volume density map, which will be utilized to derive the B-field strengths in the L1495 region based on the multi-wavelength polarisation data in the near future (Eswaraiah et al. in prep.). 

To test the~\volden~pipeline, in addition to the nearby filament L1495, we have also considered a distant filament RCW57A (also known as NGC 3576). RCW57A is an H{\sc ii} region associated with a filament and bipolar bubble and is located at a distance of 2.4 kpc (\citealt{persi_1994, Eswaraiah_2017} and references therein). It is about 30 arcmin west of the H{\sc ii} region NGC 3603 \citep{persi_1994}. The extracted number densities from \volden~are then used to re-estimate the magnetic field strengths ($B_{\mathrm{pos}}$) in various parts of RCW 57A. These results are compared with the published results by \citet{Eswaraiah_2017}.

\section{Methods}\label{methods}

\subsection{Revisiting the 3D Cylindrical Plummer Model}

Most studies observed that the column density profile follows a symmetric cylindrical geometry about the filament's spine \citep{Arzoumanian_2011}. Assuming that the filament is located on the POS, we have extracted column density profiles of the pixels along the orientation perpendicular to the spine. These profiles are fitted using the Plummer-like model to extract the model parameters ($n_{c}$, $R_{\mathrm{flat}}$ and $p$). {The model parameter $n_{c}$ represents the volume density at the filament spine, $R_{\mathrm{flat}}$ is the radius up to which $N_{p}(r)$ remains constant, and $p$ is the profile index that determines the slope of the profiles falling beyond $R_{\mathrm{flat}}$}. These parameters are then used to determine the volume densities across the radial extent of the cloud spine. Since the Plummer profile extends to infinity, a limiting cloud boundary is essential to compute the average volume density map \citep{Hoq_2017, Wang_2019}. Various studies have assumed different cloud boundary conditions for determining the mean volume density values. For instance, \citet{Hoq_2017} considered that the cloud boundary is extended up to a distance at which the column density decreases rapidly, and beyond this threshold, column density almost becomes constant. To determine this threshold boundary, they first calculated the extinction level \say{$A_{V}$} for their corresponding region by using the $N_{\mathrm{H_2}}$ to $A_V$ converting relation \citep{heiderman_etal_2010}. They reported that at an estimated extinction level of $A_{V}$ $\approx 2$ mag, this plateau in the $N_{\mathrm{H_2}}$ radial profiles occurs. Subsequently, they used the 3D Plummer model to find the average line of sight (LOS) volume densities for the corresponding $A_{V}$ contours. In contrast, \citet{Wang_2019} assumed that the central region of the filament (corresponding to the characteristic thickness of the filament) contributes half of the total column density values, and the other half contributes to the rest of the cloud extending to infinity. Based on this assumption, they defined an effective thickness of the Plummer-like profile. Finally, they calculated the mean number density along the LOS using the thickness and column density map. We have followed this approach (more details are explained in section~\ref{Proposed technique}).

\subsection{Workflow of \volden}\label{workflow_volden}

The workflow of \ volden~ can be divided into four stages: (i) input data, (ii) profile building of filament using \radfil, (iii) implementation of cloud boundary condition, (iv) obtaining the smoothed number density map. As discussed earlier in Section~\ref{contributing_data}, we have used the \textit{Herschel} column density map for the L1495 filament \citep{Palmeirim_2013}. L1495 is an elongated filament having numerous turning points along the spine, meaning that {lines perpendicular to the spine may have different position angles}, thus making it an interesting target to test our algorithm. We discuss each stage of~\volden~in more detail in the subsections below.

\subsubsection{Input Data:}

{The first component in the \volden\space workflow focuses on the input data -- which are the (i) 2D column density map and (ii) the filament spine. To find the central spine of the L1495 region (Fig.~\ref{fig1}), we have used the automated algorithm \FILFINDER~\citep{filfinder}\footnote{\href{https://github.com/e-koch/FilFinder}{Link to \FILFINDER}; {For an overview of the \FILFINDER~algorithm, including details about the input and output formats of the package, users can review the tutorial included in this repository.}}. 
We provide the 2D column density map as an input to FILFINDER. As an output, we get the 2D image containing the skeleton or spine of the filament. FILFINDER assigns boolean 1 to the pixels tracing the spine and boolean 0 to the rest of the pixels in the map. \FILFINDER~ can recover filamentary features in a wide dynamic range of brightness compared to \DISPERSE\space \citep{disperse}. The \volden\space code is designed to be flexible enough to accept a filament column density map of arbitrary resolution {as input 2D map}.}

 %Due to the ubiquitous and elongated nature of the filament structures, multiple filament finding algorithms (\FILFINDER, \DISPERSE~etc) have been developed to automate the whole process of characterizing the properties of filaments \citep{filfinder,disperse}. We have used the \FILFINDER\footnote{\href{https://github.com/e-koch/FilFinder}{Link to \FILFINDER}; {For an overview of the \FILFINDER~algorithm, the users can review the tutorial included in this repository.}} algorithm to extract the central spine of the L1495 region (Fig.~\ref{fig1}) because of its ability to recover filamentary features in a wide dynamic range of brightness compared to \DISPERSE\space \citep{filfinder}.

\subsubsection{Building radial profiles using \radfil:}

\begin{figure}%[!ht]
\centering
\includegraphics[width=9.5cm, height =8cm]{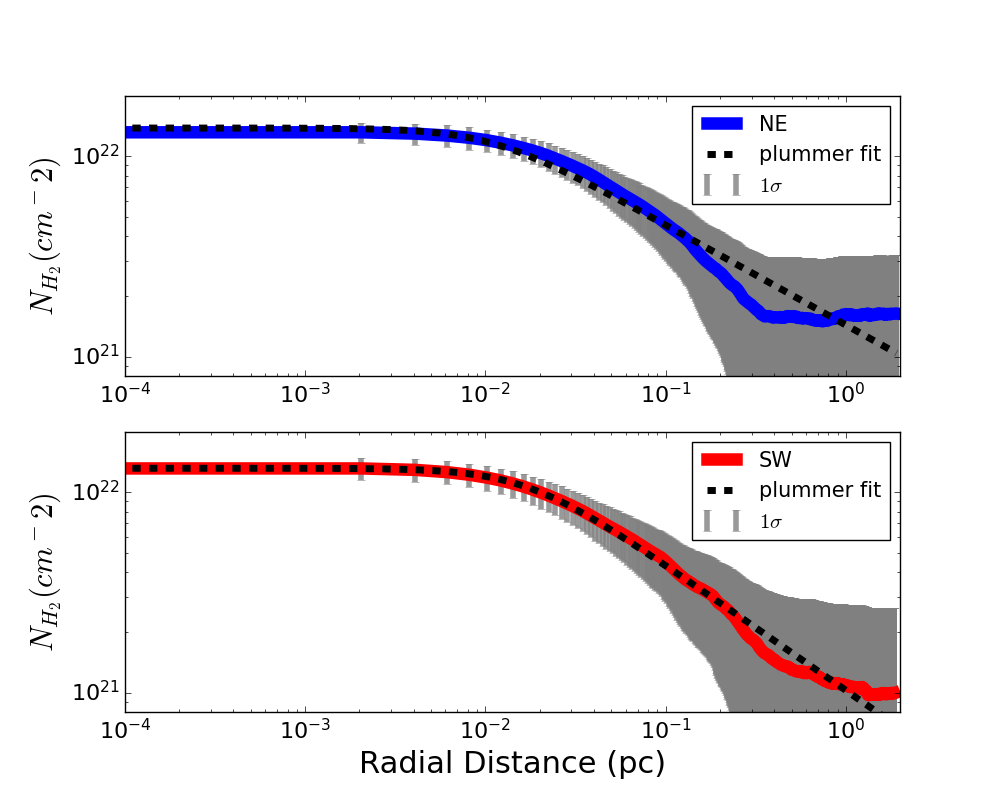}%0.6\textwidth]{taurus.eps}
\caption{The mean column density profiles of the North-East and South-West region of L1495 filament along with the Plummer-like profile fitting.}
\label{fig2}
\end{figure}

To preserve the morphology of the filament, we have used~\radfil~to build radial profiles following a \say{\textit{tailored}} approach by taking radial cuts across the filament's spine \citep{Zucker_2018}. The user can call the \textit{volden\_radfil()} function to extract the column density profiles. This is a multistep method: first, the user must select an ideal sampling frequency (\textit{sample\_int}), the distance to the cloud (\textit{distance}), and {the maximum radial distance (\textit{cut\_dist}) from the filament's spine within which data points are considered along each cut for determining the pixel showing peak intensity}\footnote{\href{https://github.com/catherinezucker/radfil}{Link to \radfil}; {To get a more illustrative view of the \textit{sample\_int} and \textit{cut\_dist} parameters, the user can follow the \radfil\space tutorial present in this repository. The user can also refer to \citet{Zucker_2018} for further details regarding the various illustrations used to describe the~\radfil~parameters and workflow.}}. Second, \radfil\space creates a smooth version of the filament spine by fitting a basis spline (B-spline) function \citep{Zucker_2018}. Finally, \radfil\space samples the B-spline points with the provided \textit{sample\_int} and creates the cuts at a constant angle \footnote{Initially, \radfil\space was designed to make cuts perpendicular to the tangent of every sampling point, eventually resulting in artifacts in the final number density map. To tackle this problem, we modified the source code of~\radfil~to extract the radial profiles perpendicular to the mean orientation of the filament's spine, thus preserving the overall cylindrical geometry of the filament.} to build the radial profiles across the spine. Figure~\ref{fig2} shows the average column density profiles. 

Here in our case, we choose the \textit{sample\_int} $=$ 6 pixels $\approx$18$\arcsec$ to match the beam width ($18.2\arcsec$) of the high-resolution column density map of the Taurus region produced by \citet{Palmeirim_2013} and $distance= 140$ pc \citep{Kenyon_etal_1994}. It is evident from Figure~\ref{fig2} that the mean $N_{\mathrm{H_2}}$ profiles start to decrease steeply beyond 0.2 pc. Therefore, we consider that the peak column density always occurs within this limit for each cut, and the $cut\_dist$ parameter in~\volden~is considered 0.2 pc.

%Thus we concluded that the peak column density will range within this limit for each cut and set $cut\_dist=$ 0.2 pc.}}

\subsubsection{Proposed technique and cloud boundary condition:}\label{Proposed technique}

In this study, we assume filament as a symmetrical cylinder across the spine, located on the POS (with an inclination angle of zero), and there are no foreground or background filaments along the direction of {L1495}. To model the column density profiles, we have used the Plummer-like model (see equation \ref{eq1} and \ref{eq2}) to fit the extracted column density profiles (e.g., \citealt{Arzoumanian_2011, Wang_2019}) 

\begin{equation}
    n_{p}(\textbf{r}) = \frac{n_{c}}{[1 + (\frac{\textbf{r}}{R_{\mathrm{flat}}})^{2}]^{\frac{p}{2}}},
    \label{eq1}
\end{equation}

\begin{equation}
    N_{p}(r) = \frac{N_{0}}{[1 + (\frac{r}{R_{\mathrm{flat}}})^{2}]^{\frac{p-1}{2}}} \, ;N_{0} = A_{p}n_{c}R_{\mathrm{flat}},
    \label{eq2}
    \end{equation}
where $N_{p}(r)$ is the 2D column density profile on the POS along a profile cut from the spine, and $n_{p}({\bf r})$ is the 3D number density at a radius ${\bf r}$ from the spine (see Figure \ref{schematic_plot} for more details). For each cut, the Plummer-like profile model yields three parameters: $R_{flat}$, $n_{c}$, and $p$, depending on the variation of the individual $N_{\mathrm{H_2}}$ profile across the spine. $A_{p}= \frac{1}{\cos{i}} \int_{-\infty}^{\infty}{ \frac{ds}{(1+s^{2})^{\frac{p}{2}}}}$ is a constant that takes account of the inclination angle made by the filament with the POS \citep{Arzoumanian_2011}. {For the assumed inclination angle of $i= 0$, the $A_{p}=\pi$. For simplicity, most studies have adopted a convention that the depths of the filamentary molecular clouds along the LOS are similar to their projected spatial extents on the POS. However, in reality, this may not be true, and hence, the assumption that the filamentary molecular clouds lie on the POS with $i= 0$ may not be valid. In that case, one can use the statistically averaged factor $A_p$ 
%resulting from the parameter $A_p$ 
to extract the more realistic number densities.
%calculate a correction factor associated with the number density values of a filament. 
Assuming that the orientation of filament is random, a statistical correction factor can be estimated by calculating the average of $\cos{\theta}$ between $0 \to \pi/2$, which is $2/\pi$. In the case of extreme inclination angles (e.g., $[0, \pi/2]$), one can estimate the average by considering different ranges (such as $\pi/4 \to 3\pi/4$).} The user can use the \textit{volden\_Plummer\_fit()} for  Plummer model fitting, which uses Scipy's non-linear least-squares fitting algorithm \textit{curve\_fit()} \citep{Virtanen_2020}.

The Plummer-like model of the cloud applies to infinity, thus describing a structure having no fixed boundaries. To tackle this issue, we select a particular spatial extent of the filament that contributes half of the total column density compared to the one over an infinite extent. This assumption helps define a characteristic thickness to compute the mean number density along each LOS. The characteristic thickness (D) along the LOS of the Plummer-like model is defined as (see \citealt{Wang_2019}) 

\begin{equation}
\frac{\int_{\frac{-D}{2}}^{\frac{D}{2}}{n_{p}(\textbf{r})dr_{los}}}{\int_{-\infty}^{\infty}{n_{p}(\textbf{r})dr_{los}}} = \frac{1}{2},
\label{eq4}
\end{equation}

\begin{figure}[hbt!]
     \centering
     \begin{subfigure}[b]{0.5\textwidth}
         \centering
         \includegraphics[width=\textwidth]{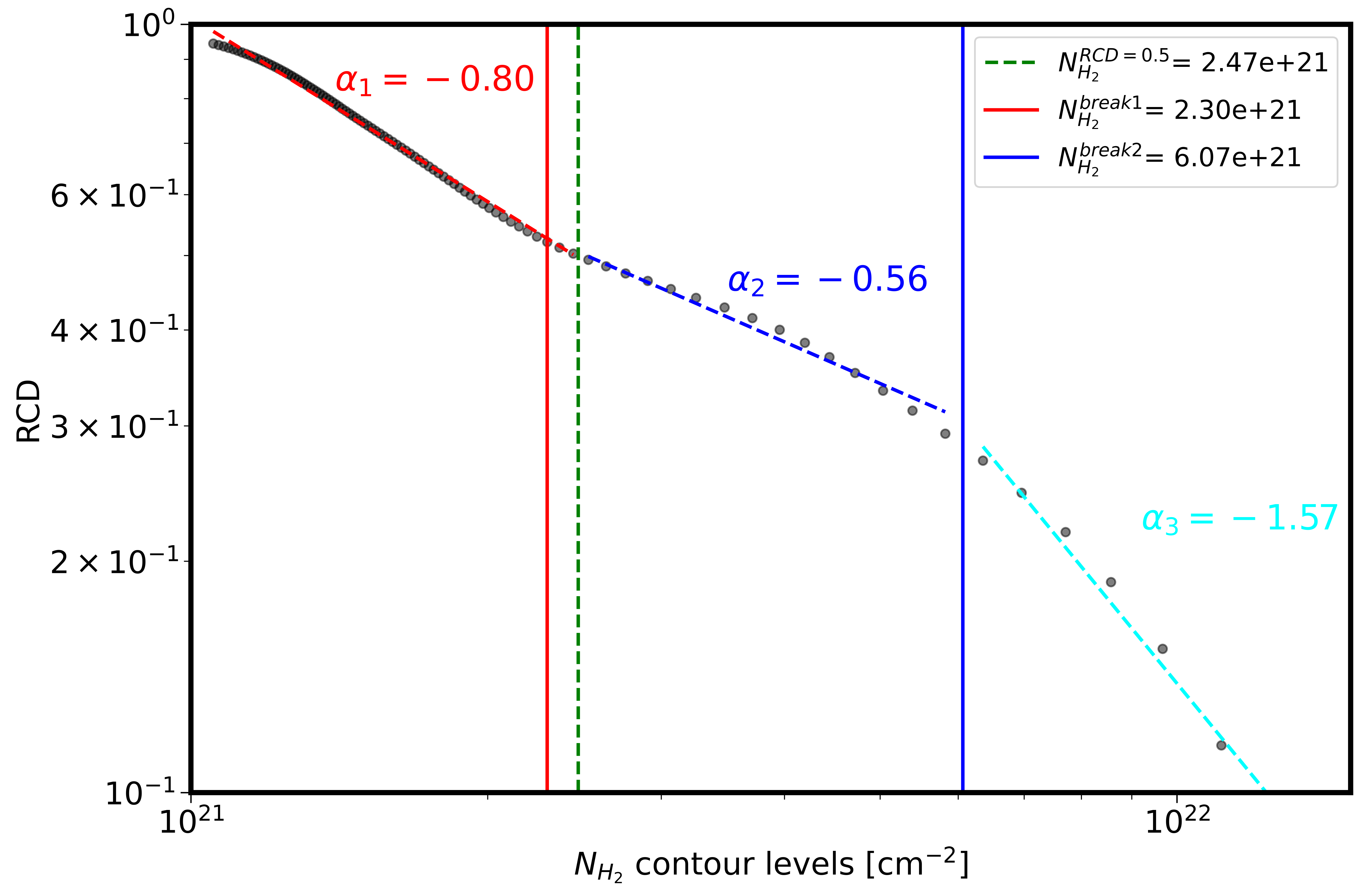}
         \caption{}
         \label{taurus_RCD_vs_contour_levels}
     \end{subfigure}
     \hfill
     \begin{subfigure}[b]{0.5\textwidth}
         %\centering
         \includegraphics[width = \textwidth]{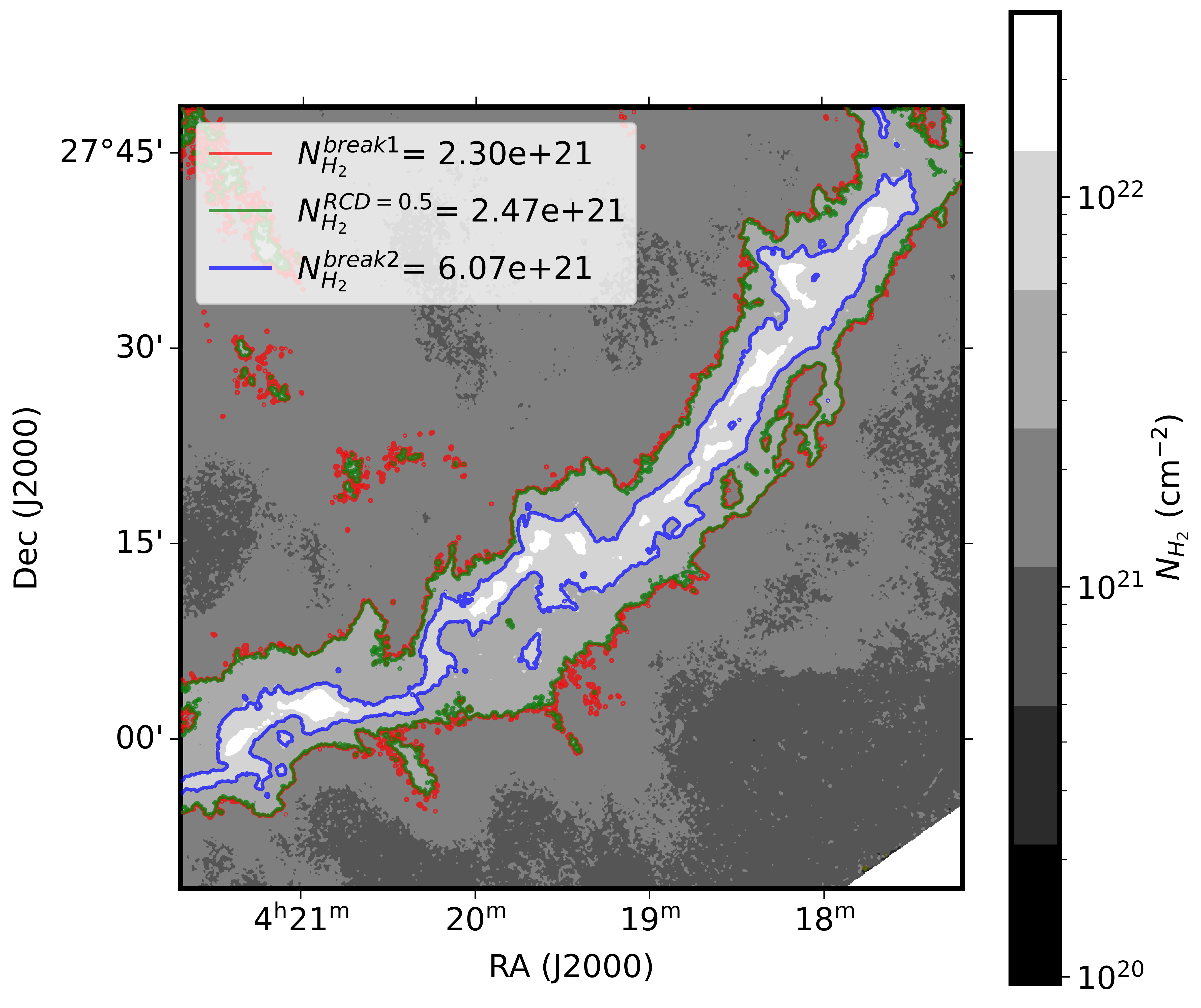}
         \caption{}
         \label{taurus_cloud_boundary_contour}
     \end{subfigure}
         \caption{(a) Column density versus RCD plot showing breaks in the power-law as indicated with red and blue vertical thick lines. (b) Herschel Gould Belt Survey  (HGBS) column density map. The red and blue contours correspond to vertical lines drawn with the same colors in {panel (a)}. {The green dotted vertical line} in panel (a) and {green contour} in panel (b) indicates the column density boundary, $N_{\mathrm{H_2}}^{\mathrm{RCD= 0.5}} \sim 2.47 \times 10^{21}$  cm$^{-2}$, within which the cloud contributes 50\% of RCD (see text for more details).}\label{cloud_boundary_explanation}
\end{figure}

%\begin{comment}
    
    \begin{figure}%[!ht]
        \centering
        \includegraphics[width=8.25cm, height =7.25cm]{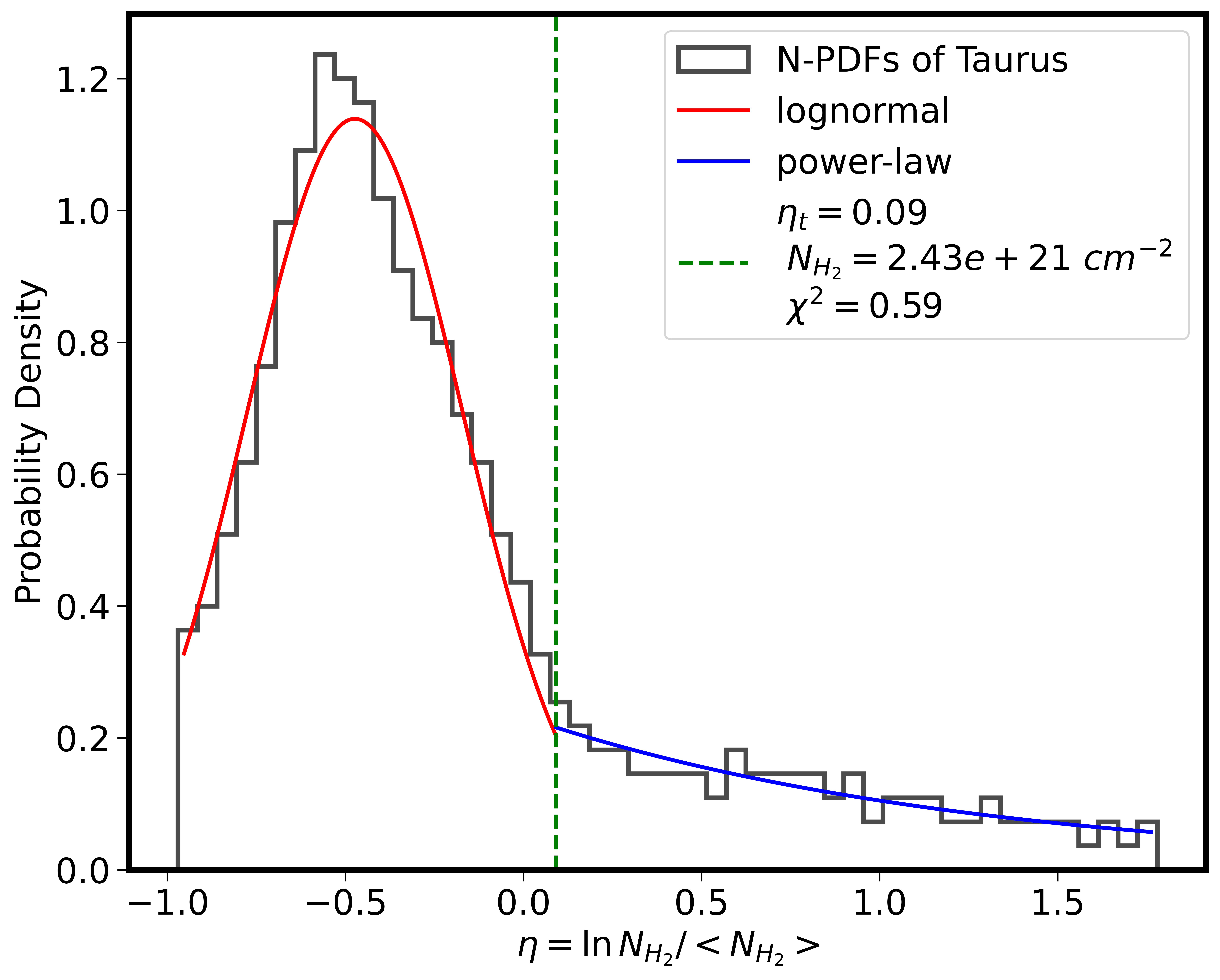}
        \caption{The N-PDF of L1495. The fitted lognormal and power-law distributions are shown using solid red and blue lines, respectively. The vertical green dotted line represents the transitional column density value derived from the $\chi^2$ analysis.}
        \label{cloud_boundary_explanation_npdfs}
    \end{figure}
%\end{comment}

where $r_{los}$ is the LOS extent over which the number density will be integrated. To test whether our assumption accurately describes the boundary of the cloud, we plotted the variation of the ratio of column densities (RCD), that is, the ratio of total column density within a particular contour level to total column density in the entire map, as a function of column density shown in Figure~\ref{taurus_RCD_vs_contour_levels}. This plot helps identify the column density level at which half of the total value arises.

We see that $50\%$ of total column density comes from the portion of the cloud lying within $N_{\mathrm{H_2}}^{\mathrm{RCD= 0.5}}=~2.47 \times 10^{21}$~cm$^{-2}$ as shown in Figure~\ref{taurus_cloud_boundary_contour}. To elucidate further, the RCD profile is fitted with the broken power laws using astropy's {\it BrokenPoweLaw1D} model; \citealt{astropy_2013}).
We evidence a change in the power-law indices of $\alpha_1 = -0.80$ to $\alpha_2 = -0.56$ at $N_{\mathrm{H_2}}^{\mathrm{break1}} = 2.30 \times 10^{21}$~cm$^{-2}$, and $\alpha_2 = -0.56$ to $\alpha_3 = -1.57$ at 
$N_{\mathrm{H_2}}^{\mathrm{break2}} = 6.70 \times 10^{21}$~cm$^{-2}$. The $N_{\mathrm{H_2}}^{\mathrm{break1}}$ closely matches with $N_{\mathrm{H_2}}^{\mathrm{RCD= 0.5}}$ as shown in the RCD and column density maps (Figure~\ref{taurus_cloud_boundary_contour}). The $N_{\mathrm{H_2}}^{\mathrm{break2}}$ encompasses the dense portion of the filament. 
%his value represents the cloud's very dense cores, and RCD rapidly decreases beyond this point due to the increased value in the column densities.

To confirm whether our assumption of 
%further cross-check whether our assumption of 
$50\%$ of the total column density represents the cloud boundary and that holds at $N_{\mathrm{H_2}}^{\mathrm{RCD= 0.5}}$, we use N-PDF analysis. This would help identify the gravitationally bounded dense regions from the low-density portion of the filament \citep{Jiao_2022}. The lognormal portion of N-PDF traces the low-density turbulent gas. In contrast, the power-law portion of the N-PDF traces self-gravitating dense molecular gas that collapses to form stars as shown in the Figure \ref{cloud_boundary_explanation_npdfs} \citep[see also][]{JKainulainenandJCTan_2013}. Therefore, NPDF can be used as another proxy to constrain the critical column density at which the N-PDF switches from lognormal to power-law\footnote{This can also be described as the cloud boundary at which gravity dominates over turbulence, and the higher density gases above this critical value collapse under gravity and form stars.} that best describes the cloud boundary. The N-PDF can be described using the notation $\eta$, and the normalized probability distribution function is given by
\begin{equation}
    \int_{\infty}^{\infty} p(\eta) d\eta = \int_{0}^{+\infty} p(N_{\mathrm{H_2}}) dN_{\mathrm{H_2}} = 1
    \label{npdf_normalised_eqn}
\end{equation}
where $\eta = \ln{N_{\mathrm{H_2}}/<N_{\mathrm{H_2}}>}$. We sampled the N-PDF by using a binned histogram approach. As mentioned earlier, the N-PDF comprises a lognormal and a power-law component. The two components are continuous at the critical (or transitional) column density \citep{JKainulainenandJCTan_2013, Chen_2018}, and the distribution is given by
\begin{equation}
    p_{\eta}(\eta) = \left\{ \begin{array}{ll} M(2\pi\sigma_{n}^2)^{-0.5}e^{-(\eta - \mu)^2/(2\sigma_n^2)} \quad \eta < \eta_t \\ M p_{0} e^{-\alpha\eta} \quad \eta > \eta_t \end{array} \right.
    \label{npdf_distribution_eqn}
\end{equation}
where $\eta_t$ is the transitional column density, $p_0$ is the amplitude of the N-PDF at the transition point, $\sigma_n$ is the dimensionless dispersion of the logarithm field, $\mu$ is the mean, $\alpha$ is the slope of the power-law tail and $M$ is the normalization factor. %Similar to \citet{JKainulainenandJCTan_2013, Chen_2018}, 
%We have carried out 
A least-square fit is performed on the N-PDF binned histograms using equation~\ref{npdf_distribution_eqn}. Subsequently, we minimized the $\chi^2$ residual value between the model (equation~\ref{npdf_distribution_eqn}) and the histogram over a range of $\eta_t$ values to calculate the optimized parameters. %Figure~\ref{cloud_boundary_explanation_npdfs} shows the results of our N-PDF analysis. From the fit, we found out 
Figure~\ref{cloud_boundary_explanation_npdfs} reveals $N_{\mathrm{H_2}}=~2.43~\times 10^{21}~\mathrm{cm^{-2}}$ to be the critical column density at which the lognormal switches to a power-law trend. This value closely matches with the column density boundary of $N_{\mathrm{H_2}}^{\mathrm{RCD= 0.5}}=~2.47 \times 10^{21}~\mathrm{cm^{-2}}$ which contributes $50\%$ of the overall column densities as clear from Figure~\ref{cloud_boundary_explanation} and Figure~\ref{cloud_boundary_explanation_npdfs}. 
%, it is evident that $50\%$ contribution of RCD arises within the column density contour of $2.47 \times~10^{21}~\mathrm{cm^{-2}}$ which is shown using green contour in Figure~\ref{taurus_cloud_boundary_contour}. 

We further used equations~\ref{eq1} and ~\ref{eq4} to compute effective thickness by numerically solving the integrals in equation~\ref{eq4} for different $r_{pos}$ as shown in Figure~\ref{schematic_plot}. To obtain the value of D, we take an initial guess value and calculate the integral number density from $-D/2$ to $+D/2$. We then calculate the ratio in equation \ref{eq4}. If the value is less-/greater than 0.5, we increase-/decrease  the guess value. As we approach the value of D at which the ratio in equation \ref{eq4} gets closer to 0.5, we stop the iterative process based on the threshold error in D defined by the user (default value is $10^{-8}$). For this purpose, the user can use the \textit{compute\_thickness\_map()} utility of~\volden~ to derive the thickness map. Finally, the mean number density values were calculated using the following relation:

\begin{equation}
%\Bar{n} = \frac{N_{H_{2}}}{2D}
{n(H_{2})} = \frac{N_{H_{2}}}{2D},
\label{eq5}
\end{equation}

where $N_{H_{2}}$ is the observed column density at a given pixel and $D$ is the computed effective thickness. The number density map can be calculated using~\volden's \textit{number\_density\_map()} method.

%\subsubsection{Smoothing:}\label{smooth}

%The density distribution within the filament may not precisely match with the Plummer-like model owing to the inhomogeneous structures in the column density maps, which result in low-level inhomogeneities in the produced thickness and number density maps. To reduce those inhomogeneities, the maps should optionally be smoothed. We have used astropy's \textit{Gaussian2DKernel} as a filter \citep{astropy_2013}. We choose the standard deviation in both $x$ and $y$ dimensions of the kernel to be $\sim2.58$ pixels, which is eight times the standard deviation ($\sigma=$ \textit{sample\_int}/2.355 pixels).
%Since the image dimension and kernel size were large, we have used the fast Fourier transform (FFT) as a convolution technique to save computation time \citep{astropy_2013}. 

\begin{figure}%[!ht]
\centering
\includegraphics[width=8.25cm, height =7.25cm]{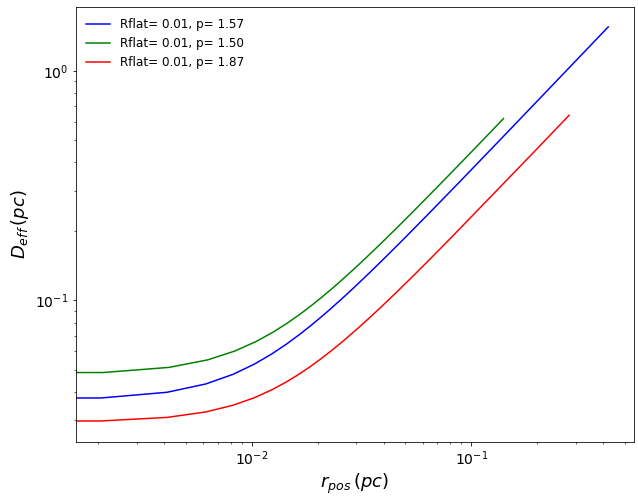}
\caption{Variation of the computed effective thickness ($D_{\mathrm{eff}}$) versus radial offset in the plane of the sky ($r_{\mathrm{pos}}$). For $r_{\mathrm{pos}} >> R_{\mathrm{flat}}$, $D_{\mathrm{eff}}$ varies almost linearly with $r_{\mathrm{pos}}$.}
\label{fig3}
\end{figure}

\begin{figure*}
\centering
\includegraphics[width=\textwidth]{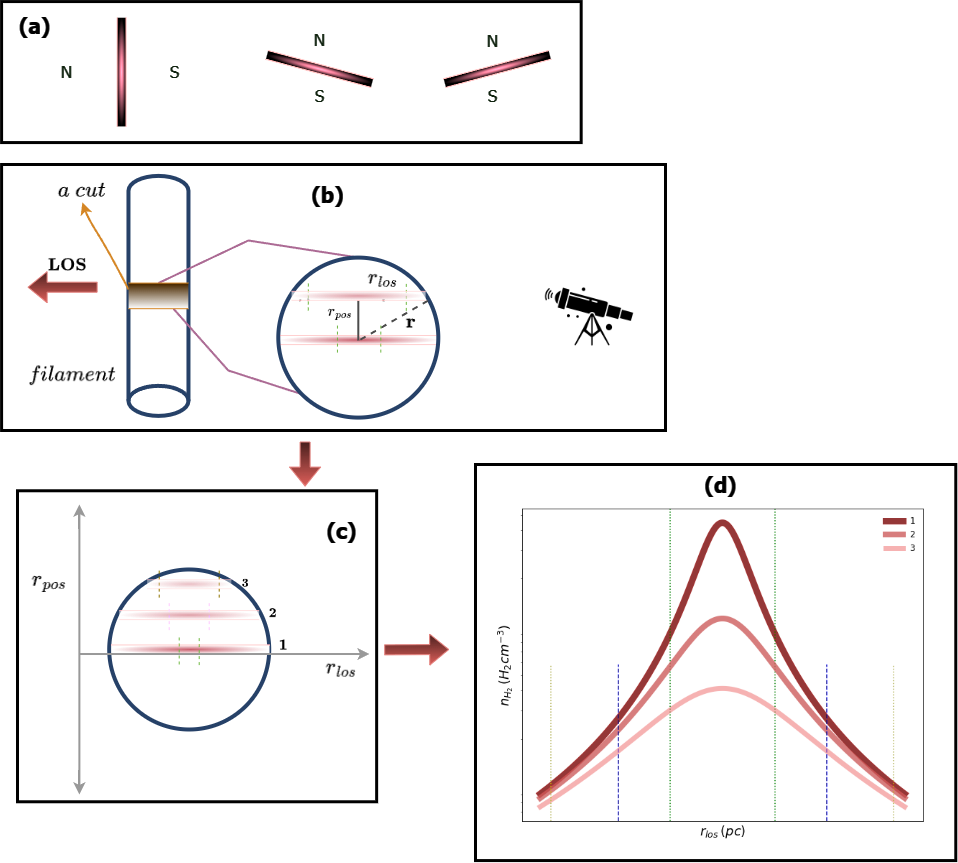}
\caption{Schematic view of the workflow involved in $\volden$. (a) The convention in which the direction of the filament is divided into north (N) and south (S) in the $\volden$ workflow. For simplicity purposes, we have used the same N and S convention when the overall orientation of the filament is perpendicular to the horizontal axis. However, the user can convert it into east (E), west (W), or any other direction(s), depending on their filament's orientation. (b) Visualizing edge-on-view and face-on-view of a cut along the filament's minor axis. The relation between $r_{los}$, $r_{\mathrm{pos}}$, and the 3D radius (\textbf{r}) from the central spine according to the Plummer-like profile model is such that $\textbf{r} = \sqrt{r_{\mathrm{pos}}^2 + r_{los}^2}$. (c) Visualization of three arbitrary cuts along LOS (LOS-cuts). The differences in the spatial extents of colors along the LOS cuts are attributed to the fact that the gas is more compact along the spine and diffused in the outer extent of the cloud. {The vertical dashed lines represent the characteristic thickness of the filament at a particular~$r_{\mathrm{pos}}$}. Note that $r_{\mathrm{pos}}$ = 0 represents the spine points along the filament's central skeleton. (d) The expected number density profiles along the LOS-cuts 1, 2, and 3. As evident, LOS-cut 1 exhibits higher density but is confined to the central portion of the filament. In contrast, the LOS-cut 3, distributed over a larger extent, shows lower-peak density (marked as yellow dashed vertical lines).}
\label{schematic_plot}
\end{figure*}

\section{Results}

\subsection{Application of~\volden~to L1495}\label{taurus_results}

As mentioned in section~\ref{Proposed technique}, the $N_{\mathrm{H_{2}}}$ profiles are fitted with the Plummer-like model (equation~\ref{eq2}) to extract the model parameters. The resultant mean fitted values $n_{c}$, $R_{\mathrm{flat}}$ and $p$ of the North-West part of L1495 cloud are found to be ($8 \pm 6$) $\times10^4$ cm$^{-3}$, $0.033 \pm 0.030$ pc and $1.76 \pm 0.27$, respectively. Similarly, for the South-East part of L1495 cloud, $n_{c}$, $R_{\mathrm{flat}}$ and $p$ are found to be ($7 \pm 6$) $\times10^4 cm^{-3}$, $0.036 \pm 0.028$ pc and $1.85 \pm 0.25$ respectively. These results agree with the model parameters obtained by \citet{Palmeirim_2013} for the L1495 segment ($p$ = 1.7 and $R_{\mathrm{flat}}$ = {0.02 pc}). The fitted values of $n_{c}$, $R_{\mathrm{flat}}$ and $p$ for the entire region are in the range of ($\sim$0.6~--~30)$\times$10$^4$ cm$^{-3}$, 0.010~--~0.090 pc and 1.32~--~2.23 for all the profiles of the cloud spine.

\begin{figure}%[!ht]
\centering
\includegraphics[width=8.25cm, height =7.25cm]{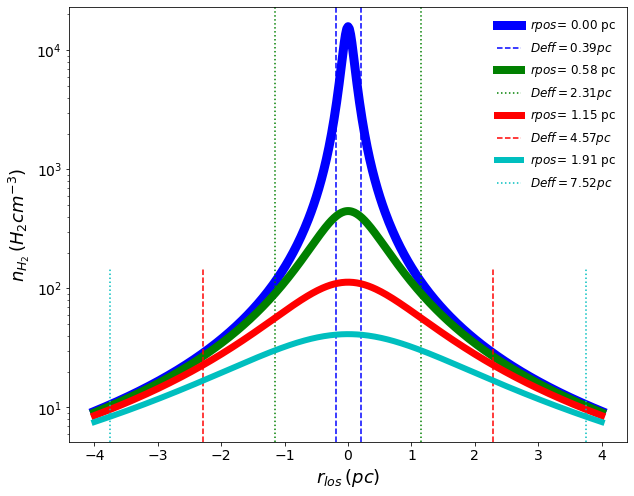}
\caption{Variation of number density profiles $n(H_{2})$, as a function of LOS distance from the spine (see panel (d) of Figure \ref{schematic_plot}). Profiles along the filament ridge show higher densities and a higher degree of compactness than those in the outer portions of the filament. As a result, a reduced thickness is witnessed throughout the central ridge of the filament (e.g., blue vertical lines). In contrast, a larger thickness is observed in the outer portions of the filament (e.g., cyan vertical lines).}
\label{rlos_vs_nh2}
\end{figure}

Figure~\ref{fig3} shows the variation of computed effective thickness values versus the POS radial offset from the spine, which suggests that at higher $r_{\mathrm{pos}}$ ($r_{\mathrm{pos}} >> R_{\mathrm{flat}}$), the thickness values varies almost linearly with $r_{\mathrm{pos}}$. At $r_{\mathrm{pos}}$ = 0, one would naturally expect higher thickness at the filament's ridge, but we determine it to be much lower because of the compact filament ridge, as evident from Figures~\ref{schematic_plot} and \ref{rlos_vs_nh2}. 

\begin{figure*}
\centering
\begin{multicols}{2}
    \includegraphics[width=\linewidth]{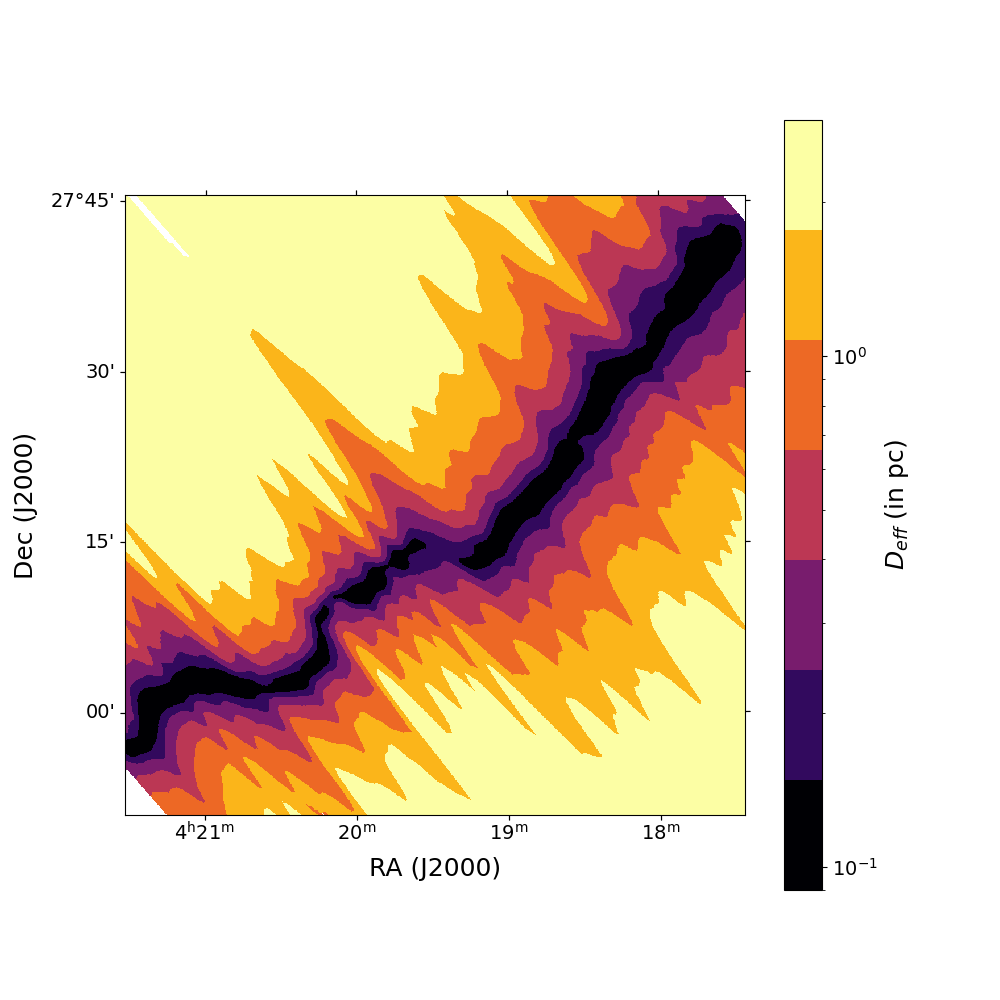}\\(a)

    \includegraphics[width=\linewidth]{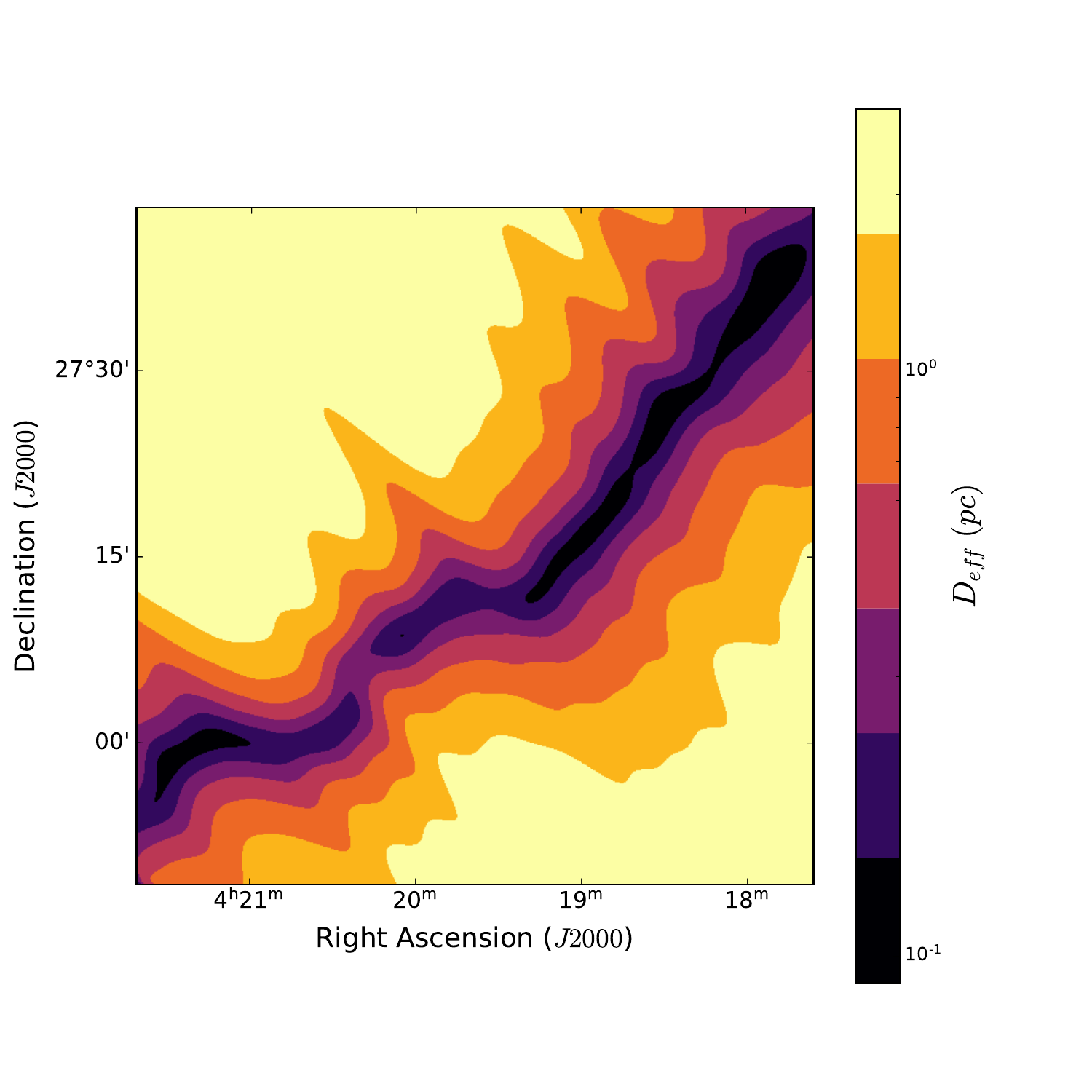}\\(b)
    
\end{multicols}
\caption{ \textcolor{blue}{(a)} and \textcolor{blue}{(b)}: The thickness map (before and after smoothing) for L1495 obtained by using the \textit{compute\_thickness\_map()} method. This method uses equation~\ref{eq1} and equation~\ref{eq4} to derive the thickness values of the molecular cloud numerically.}
\label{fig5}
\end{figure*}

\begin{figure}
\includegraphics[width=9.25cm, height =9.25cm]{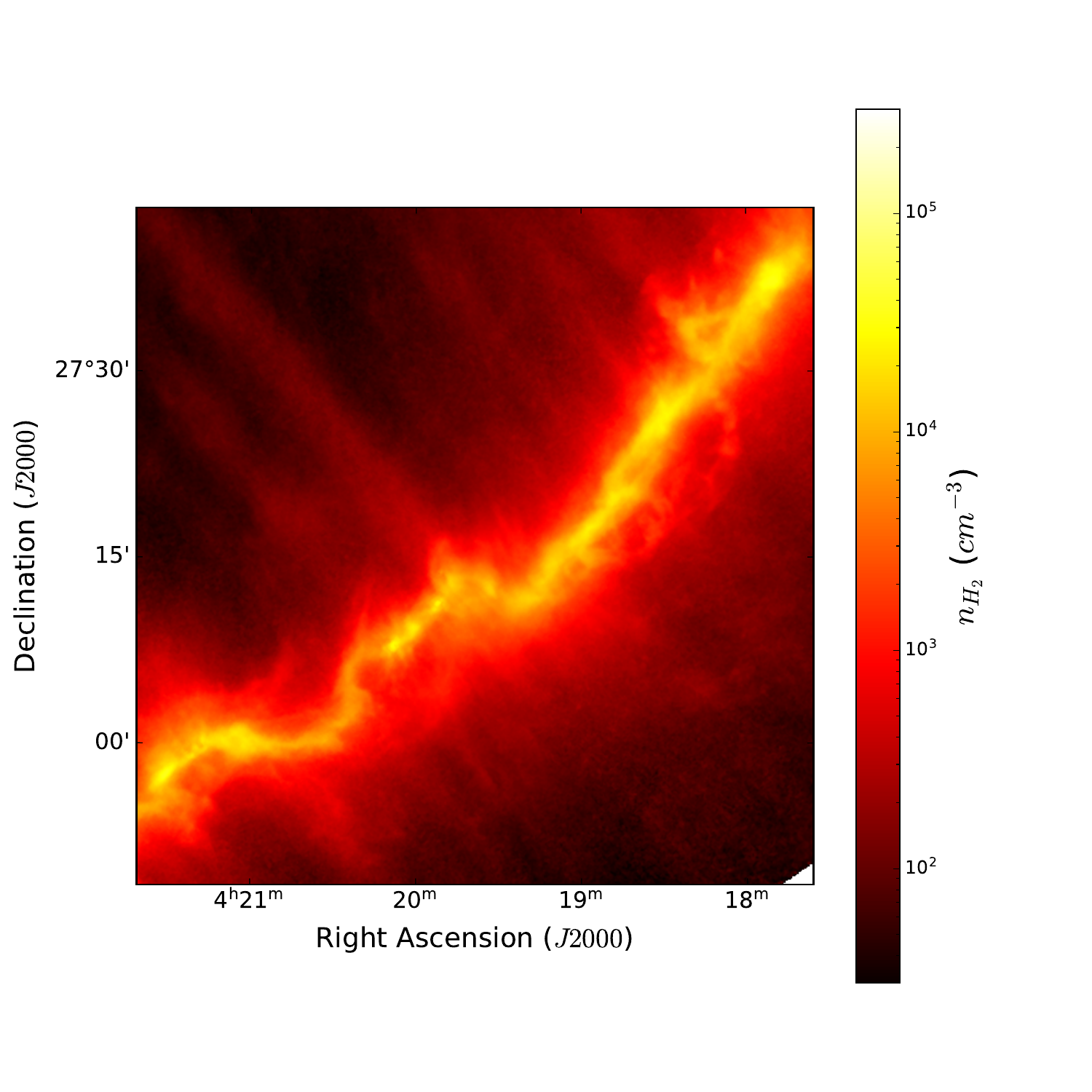}%0.6\textwidth]{taurus.eps}
\caption{The number density map obtained from the column density and thickness map by employing~\volden.}
\label{fig6}
\end{figure}

%A smooth version of the thickness map of the L1495 region was obtained by using astropy's \textit{Gaussian2DKernel} with a kernel size of $60.67\arcsecond \times 60.67\arcsecond$ \citep{astropy_2013}. 
%A smooth version of the thickness map of the L1495 region was obtained by using the methods described in section~\ref{smooth}. Figure~\ref{fig5} presents the smoothed thickness map derived from the column density maps. Using equation \ref{eq5}, the volume density map is produced by dividing the column density map with two times the smoothed thickness map and is shown in Figure~\ref{fig6}.
A smooth version of the thickness map of the L1495 region was obtained by using astropy's \textit{Gaussian2DKernel} \citep{astropy_2013} with a kernel size of $20.64~\text{pix}~\times~20.64~\text{pix}$\footnote{{The {\textit{Gaussian2DKernel}} function in \textit{Astropy} requires two parameters: \textit{x\_stddev} and \textit{y\_stddev}, which represent the standard deviations $\sigma$ of the Gaussian kernel in the x and y directions. Using the relation $\sigma = \text{FWHM} / 2.355$, we calculated $\sigma = 2.58~\text{pix}$. We initially performed the convolution with this $\sigma$ value and observed some artifacts due to missing pixel values in the resulting thickness map. We used a trial-and-error approach to determine the optimal kernel size to address these artifacts and ensure that the convolution appropriately fills in any missing pixels. We found that setting the kernel size to $8\sigma \times 8\sigma$ ($20.64~\text{pix} \times 20.64~\text{pix}$) minimized these artifacts.}}. Figure~\ref{fig5} presents the thickness map (before and after smoothing) derived from the column density maps. Using equation \ref{eq5}, the number density map is produced by dividing the column density map by two times the smoothed thickness map and is shown in Figure~\ref{fig6}.

\begin{comment}
    
\begin{figure}
\centering
\includegraphics[width=9.25cm, height =9.25cm]{taurus_plots/taurus_thickness_map.pdf}
\caption{The thickness map for L1495 obtained by using the \textit{compute\_thickness\_map()} method. This method uses equation~\ref{eq1} and equation~\ref{eq4} to derive the thickness values of the molecular cloud numerically.}
\label{fig5}
\end{figure}
\end{comment}

%To check the effectiveness of smoothing, we have first produced the histogram distribution of unsmoothed versus smoothed thickness maps in Figure~\ref{log_log_plot_taurus}. Furthermore, we reproduced the column density map using the thickness and volume density map by rearranging the terms in equation~\ref{eq5} and derived an offset map (Figure~\ref{offset_map}) to compare it with the original column density map produced by \citet{Palmeirim_2013}. In Figure~\ref{offset_histogram}, we plotted the histogram distribution of the produced offset map, which suggests a symmetric distribution centered at zero. Therefore, we conclude that the assumed Plummer-like profile model is capable of modeling the column density map and that the smoothing has a minor or negligible effect on the resulting number density map.

\begin{figure*}
\centering
\begin{multicols}{2}
    \includegraphics[width=\linewidth]{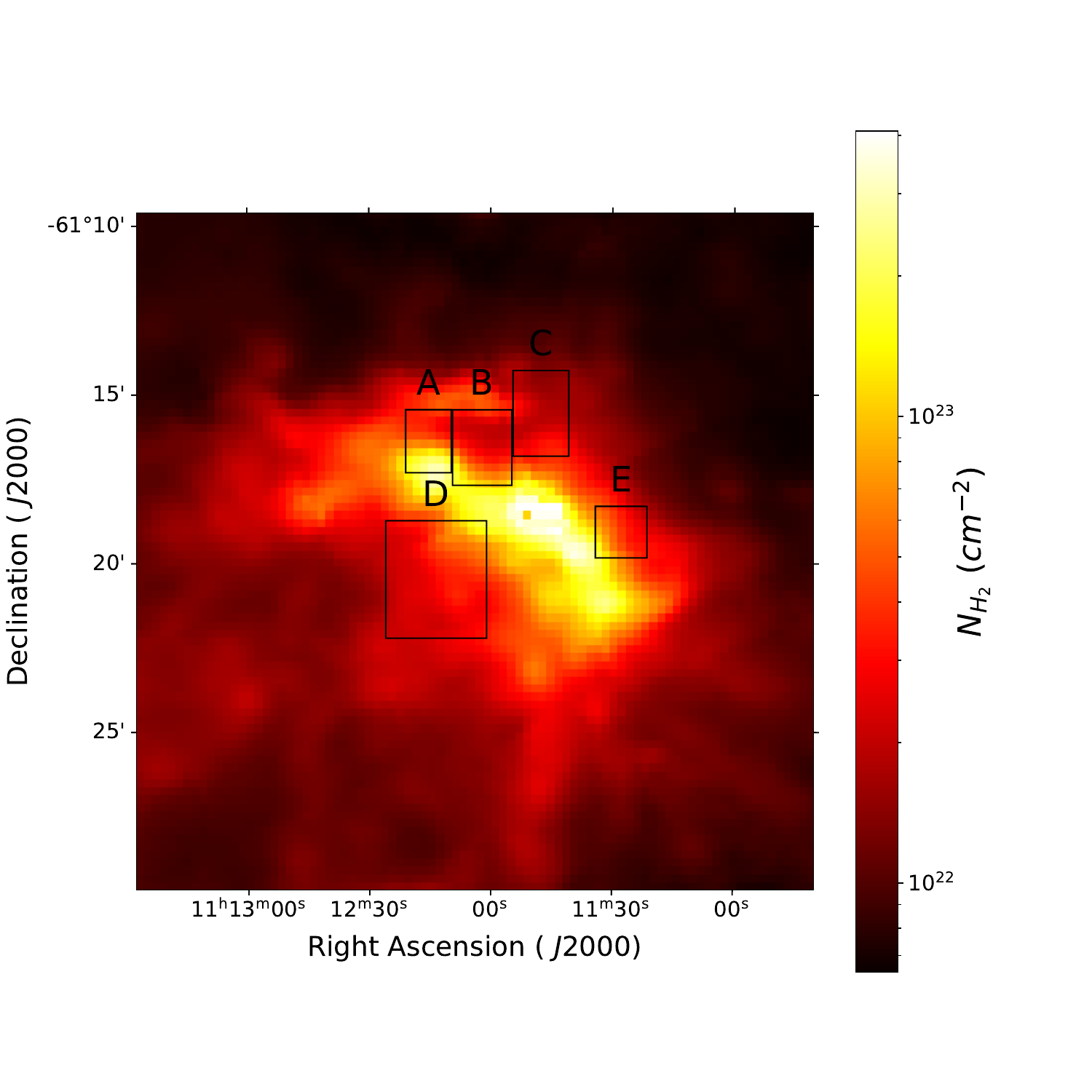}\\(a)

    \includegraphics[width=\linewidth]{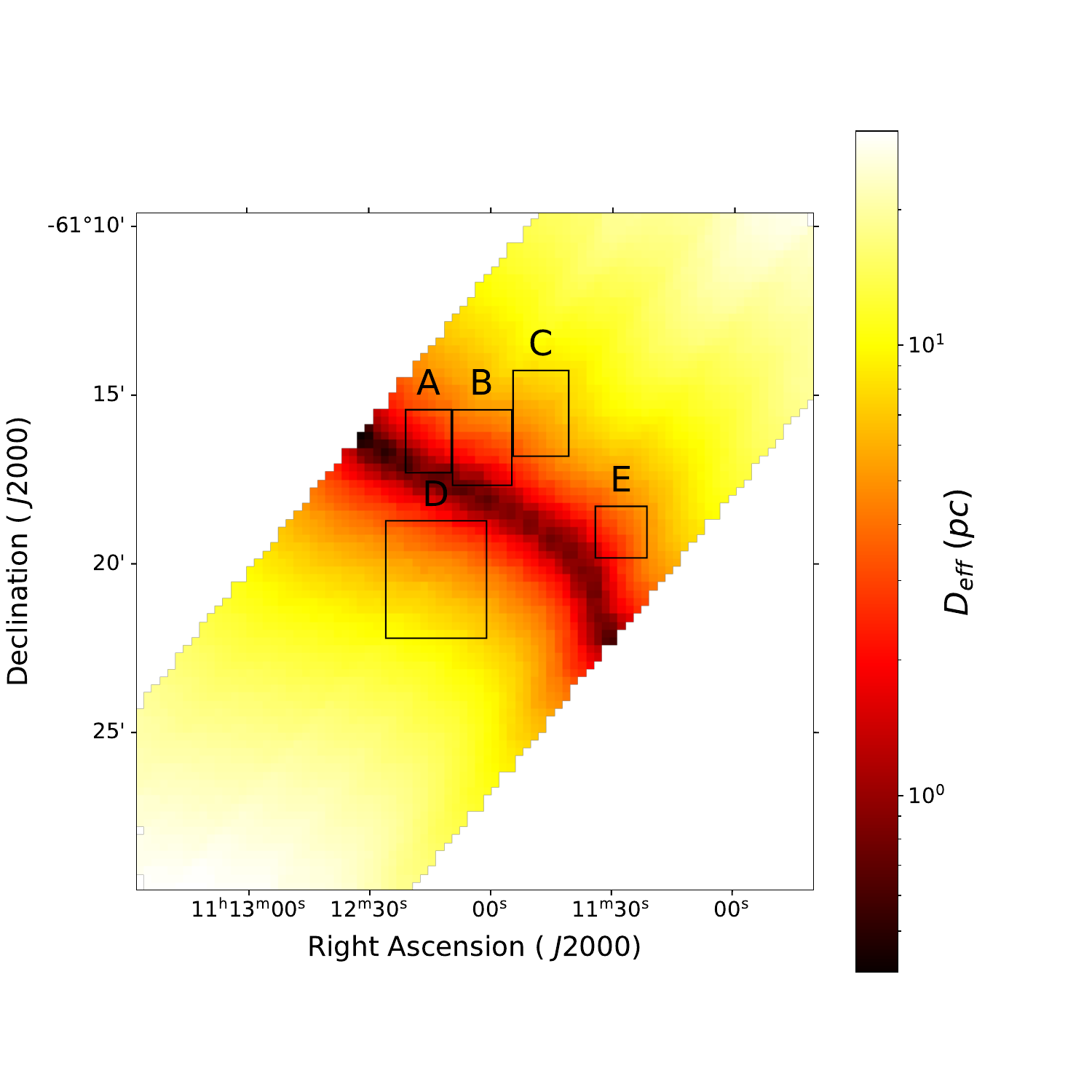}\\(b)
    
\end{multicols}
    
\begin{multicols}{2}
    \includegraphics[width=\linewidth]{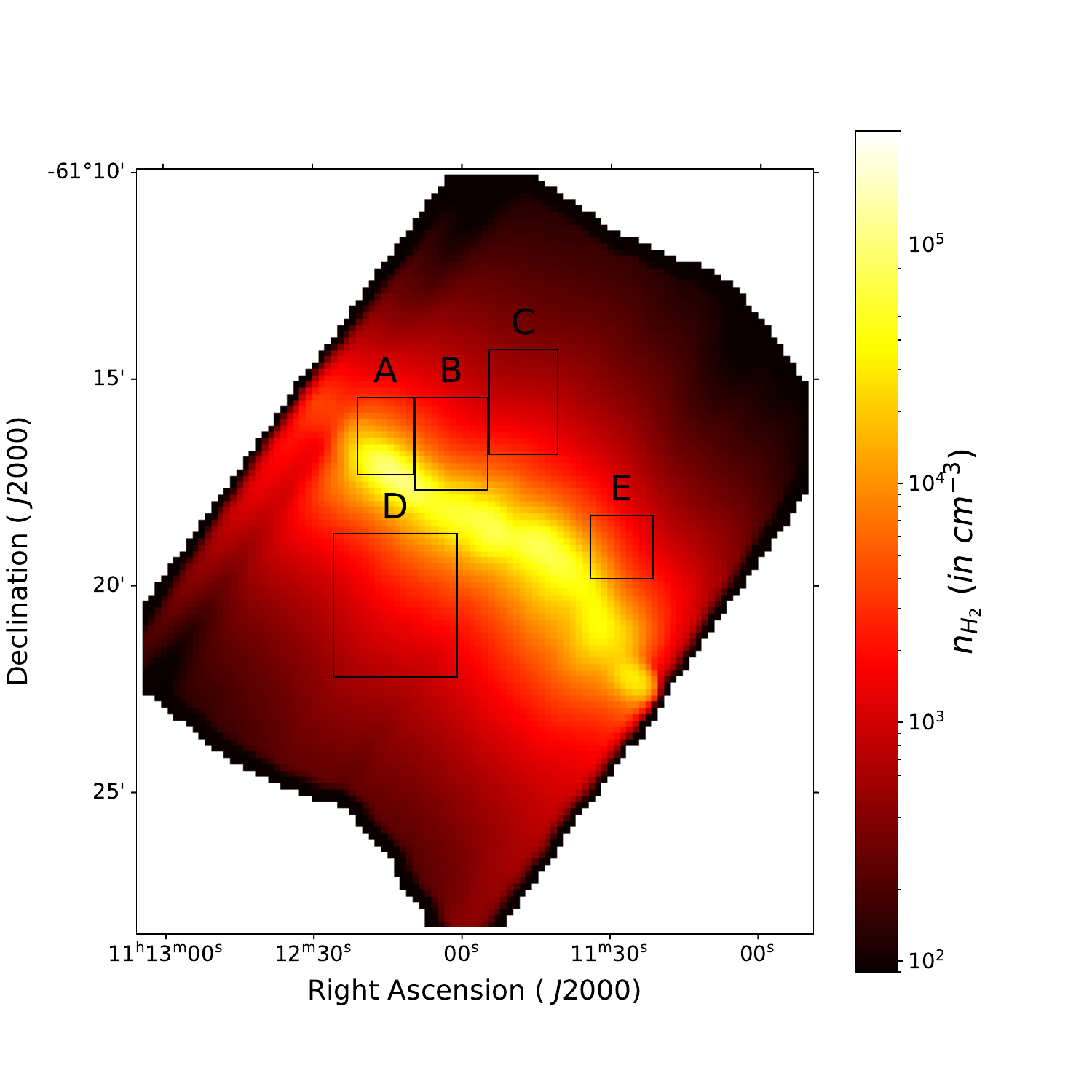}\\(c)

    \includegraphics[width=\linewidth]{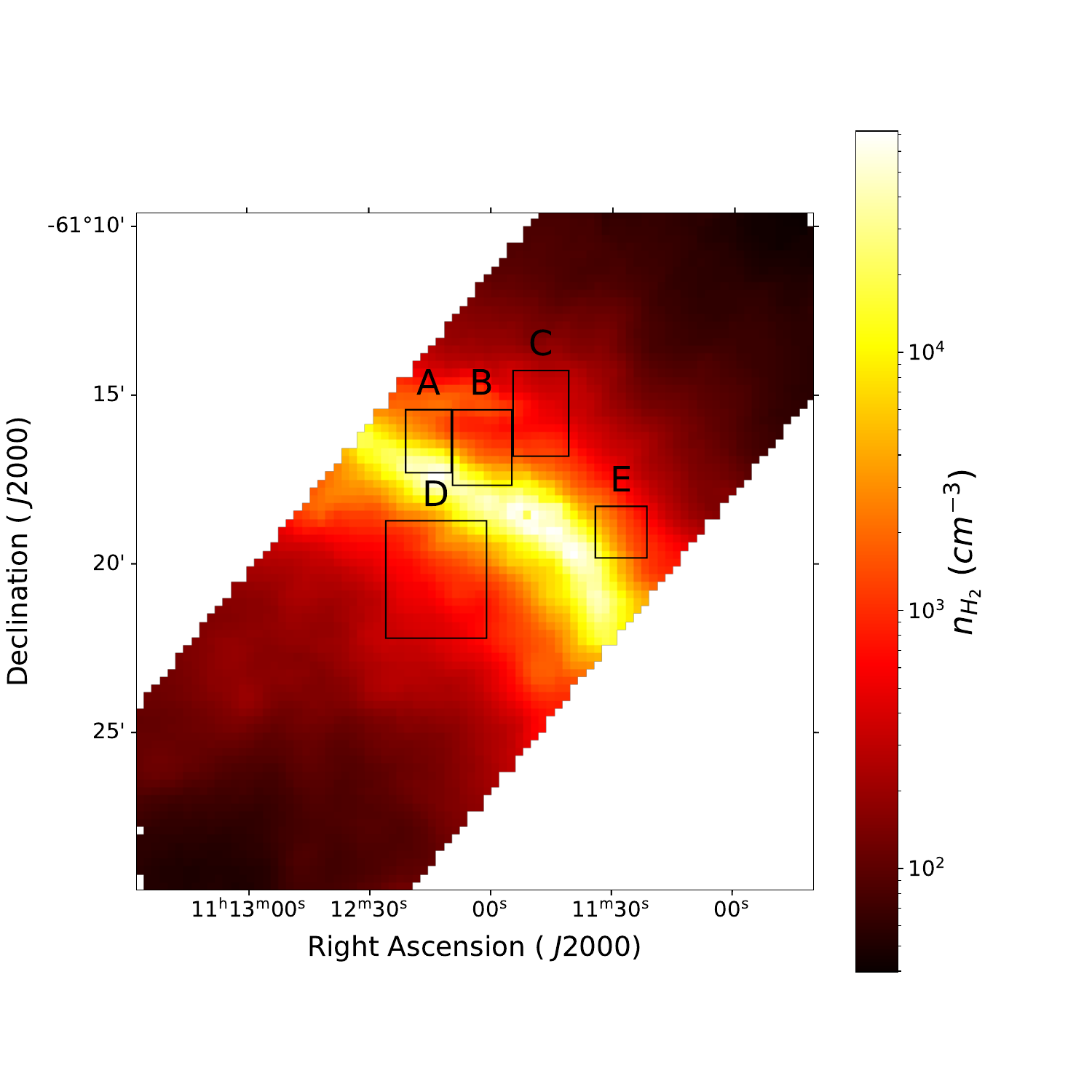}\\(d)
    
\end{multicols}
\caption{ \textcolor{blue}{(a)} and \textcolor{blue}{(c)}: The Column density and number density map of RCW 57A obtained by \citet{Eswaraiah_2017}. \textcolor{blue}{(b)} and \textcolor{blue}{(d)}: The thickness and number density map obtained through the~\volden~pipeline.}
\label{rcw_results}
\end{figure*}

The resulting number density map of the L1495 region, along with the multi-wavelength polarization data, will be used to perform the DCF analyses to obtain a coherent picture of how the magnetic field strength, mass-to-flux ratio criticality, and Alfv\'{e}nic Mach number vary as a function of column density and spatial scales (Eswaraiah et al. in prep). 

\subsection{Application of~\volden~to RCW 57A}\label{concept_proof}

To test the reliability of~\volden, we derived the number density map of RCW 57A using the described method. We compared our results with the number density results published by \citet{Eswaraiah_2017}. The distance to the filament was considered to be 2.4 kpc \citep{Eswaraiah_2017}. To produce the volume density map, we have adopted the beamwidth of 36.4$\arcsec$, the same as that of the column density map obtained by \citet{Eswaraiah_2017}. While constructing profiles using~\radfil, we have assumed a profile sampling of \textit{sample\_int}= 3 pixels to match the beamwidth of 36.4$\arcsec$. We found that more than 95\% of the polarization vectors are centered outside of the contour corresponding to the volume density $n(H_{2}) = 10^4$ cm$^{-3}$. It implies that the NIR polarimetry fails to adequately trace the highly extincted parts (\citet{Eswaraiah_2017} and the references therein). Therefore, we exclude the pixels with $n \geq 10^4$ cm$^{-3}$ while estimating the mean volume densities in regions\footnote{{\citet{Eswaraiah_2017}  have observed the B-field structure in the entire region of RCW57A is quite complex; however, in each specific portion of the region, the B-field orientations appear to be organized with a specific mean B-field orientation in comparison to the other region. We followed their approach and divided the observed area of RCW 57A into five different regions, named A, B, C, D, and E, as shown in Figure~\ref{rcw_results}.}} A, B, C, D, and E of RCW 57A (see Figure~\ref{rcw_results}). We then used {velocity dispersion} ($\sigma_{V_{LSR}}$) and {dispersion in polarization angles} ($\sigma_{\theta(H)}$) values from \citet{Eswaraiah_2017} to calculate the B-field strengths using the DCF \citep{Chandrasekhar_fermi} as well as the ST \citep{Skalidis_Raphael_2021} methods. These results are summarized in Tables~\ref{table1} and \ref{table2}.

By considering the uncertainties, the Plummer model parameters produce results that closely agree with \citet{Eswaraiah_2017}, thus showing that~\volden~is a good implementation of the Plummer fit.
Figure~\ref{rcw_results}\textit{\textcolor{blue}{(c)}} and \ref{rcw_results}\textit{\textcolor{blue}{(d)}} shows the volume density produced by \citet{Eswaraiah_2017} and~\volden~respectively. From Table~\ref{table2}, we concluded that both $n(H_{2})$ and B-field values in the case of~\volden~{are smaller} by a factor of $\approx 1.4$ as compared to \citet{Eswaraiah_2017}. {This difference is due relatively higher number densities} obtained by \citet{Eswaraiah_2017}. They treated the Plummer-like profile model as 2D and did not employ the cloud boundary condition.

%table 1
\begin{table*}
\centering
\begin{tabular}{lcccccl}\toprule
& \multicolumn{3}{c}{\citet{Eswaraiah_2017}} & \multicolumn{3}{c}{\volden}
\\\cmidrule(lr){2-4}\cmidrule(lr){5-7}
           Region & \thead{$n_{c}$\\$(\times~10^4~cm^{-3})$} & $R_{\mathrm{flat}}~(pc)$ & $p$    & \thead{$n_{c}$\\$(\times~10^4~cm^{-3})$} & $R_{\mathrm{flat}}~(pc)$ & $p$\\\midrule
NW    & $7.6 \pm 1.3$ & $0.2 \pm 0.1$ & $2.1 \pm 0.1$ & $4.3 \pm 0.9 $ & $0.24 \pm 0.03$ & $1.93 \pm 0.01$ \\
SE    & $5.1 \pm 0.9$ & $0.4 \pm 0.2$ & $2.0 \pm 0.1$ & $4.5 \pm 0.7$ & $0.23 \pm 0.03$ & $1.93 \pm 0.01$ \\\bottomrule
\end{tabular}
\caption{Comparison of the calculated Plummer model parameters for RCW 57A obtained using~\volden~with that of \citet{Eswaraiah_2017}.}
\label{table1}
\end{table*}

% Table 2: not-including info about dispersion in polarization angles and velocity dispersion

\begin{table*}
\centering
\begin{tabular}{lcccccl}\toprule
& \multicolumn{2}{c}{\citet{Eswaraiah_2017}} & \multicolumn{4}{c}{\volden}\\\cmidrule(lr){2-4}\cmidrule(lr){5-7}
           Region & \thead{$n_{c}$\\$(\times~10^3$ cm$^{-3}$)} & \thead{$B$-field (DCF)\\($\mu G$)} & \thead{$B$-field (ST)\\($\mu G$)}  &  \thead{$n_{c}$\\$(\times~10^3$ cm$^{-3}$)} & \thead{$B$-field (DCF)\\($\mu G$)} & \thead{$B$-field (ST)\\($\mu G$)}\\\midrule
A    & $5.15 \pm 0.68$ & $123 \pm 89$ & $73 \pm 28$ & $2.8 \pm 0.3 $ & $91 \pm 5 $ & $54 \pm 3 $ \\
B    & $3.51 \pm 0.41$ & $89 \pm 21$ & $67 \pm 11$ & $1.9 \pm 0.2 $ & $70 \pm 3$ & $50 \pm 3 $\\
C    & $1.01 \pm 0.09$ & $63 \pm 23$ & $39 \pm 11$ & $0.4 \pm 0.02$ & $43 \pm 1$ & $27 \pm 1 $\\
D    & $2.52 \pm 0.21$ & $108 \pm 36$ & $75 \pm 21$ & $1.6 \pm 0.1 $ & $87 \pm 4$ & $61 \pm 2 $\\
E    & $3.33 \pm 0.52$ & $74\pm 50$ & $53 \pm 19$ & $1.6 \pm 0.1$ & $51 \pm 3 $ & $37 \pm 2 $\\\hline
\end{tabular}
\caption{Comparison of number density values and magnetic field strength of RCW 57A obtained by \citet{Eswaraiah_2017} and~\volden~for region A, B, C, D, and E. The $\sigma_{V_{LSR}}$ and $\sigma_{\theta(H)}$ values for RCW 57A are considered to be the same as that of Table 2 from \citet{Eswaraiah_2017}. We have used both DCF \citep{Chandrasekhar_fermi} and ST \citep{Skalidis_Raphael_2021} methods to calculate the B-field strengths.}
\label{table2}
\end{table*}

\section{Conclusions}

We have developed a python tool~\say{\volden}~to extract the number density map from the column density maps of elongated filamentary structures in the ISM.~\volden~uses the workflow of~\radfil~\citep{Zucker_2018} to extract the radial column density profiles of the filament by assuming the geometry to be cylindrically symmetric.~\volden~then uses the Plummer-like model to extract the various fitting parameters and subsequently utilizes a boundary condition (\citealt{Wang_2019}) to derive the thickness map. Finally, the number density map is produced using the column density map and the thickness map. We chose L1495 as our primary target in this study due to its elongated and curved filament morphology. We find out that the Plummer model parameters agree with that of \citet{Palmeirim_2013}. {Through the N-PDF analysis, we determine that our cloud boundary condition is accurate and~\volden~can efficiently produce the number density map of the curved filament L1495.} To further test our analysis, we have extracted the number density map of RCW 57A and compared the number density map and B-field values with the published study by \citet{Eswaraiah_2017}. We find that the Plummer parameters ($R_{\mathrm{flat}}$ and $p$) closely agree with the values obtained by \citet{Eswaraiah_2017}. However, the B-field and $n(H_{2})$ values obtained in~\citet{Eswaraiah_2017}~are higher by a factor of $\approx 1.4$ times the values obtained in the case of~\volden. The discrepancy can be accounted for by the fact that no cloud boundary condition was employed, and the Plummer-like
profile model was considered to be 2D in \citet{Eswaraiah_2017}.

\volden~will be available at GitHub\footnote{\href{https://github.com/aa16oaslak/volden}{Link to~\volden}} upon publication. Users can follow the Tutorial in the provided GitHub link to get acquainted with the workflow and utilities of~\volden~. Users are welcome to clone the GitHub repository and make changes if required as ~\volden~is open source. We urge you to report any problems or feature requests to the authors. \volden ~can be run in Python 3.5, Python 3.6, or Python 3.9 (Python 3.9.12 is recommended). 

Currently,~\volden~can only extract the number density map of elongated filaments with the central spine, similar to Taurus L1495 ({e.g., Musca, Pipe nebula}). As long as the filament's central spine is somewhat straight and does not contain sub-filaments/branches,~\volden~can be used effectively. In the future versions of~\volden, we hope to expand \volden's functionality so that it can be applied to filaments having branched morphologies. We envision that~\volden~would be a genuine extension of existing filament identification and radial profiles packages by providing additional functionality and flexibility for the users interested in doing a robust study of POS B-field strengths in the star-forming filaments.

\section*{Acknowledgements}

%This work has been carried out as a part of the summer and semester projects under the supervision of CE during 1/2022 to 11/2022 at IISER Tirupati.  
{We thank the anonymous referee for thoroughly reviewing the manuscript and providing constructive comments, which have improved the manuscript's contents.}
This work has been carried out as a part of the summer and semester projects from January to November 2022 under the supervision of EC at IISER Tirupati. EC acknowledges the financial support from grant RJF/2020/000071 as a part of the Ramanujan Fellowship awarded by the Science and Engineering Research Board (SERB), Department of Science and Technology (DST), Govt. of India. We want to thank the authors of \citet{Zucker_2018} for releasing the~\radfil~package publicly available for the community. This research has made use of data from the Herschel Gould Belt survey (HGBS) project (\url{http://gouldbelt-herschel.cea.fr}). The HGBS is a Herschel Key Programme jointly carried out by SPIRE Specialist Astronomy Group 3 (SAG 3), scientists of several institutes in the PACS Consortium (CEA Saclay, INAF-IFSI Rome, and INAF-Arcetri, KU Leuven, MPIA Heidelberg), and scientists of the Herschel Science Center (HSC).

%%%%%%%%%%%%%%%%%%%%%%%%%%%%%%%%%%%%%%%%%%%%%%%%%%
\section*{Data Availability}

The column density data of Taurus L1495 is available in the Herschel Gould Belt survey (HGBS) data archive (\url{http://gouldbelt-herschel.cea.fr}). {The various maps, results obtained,~\volden~tutorial, and the codes used for the~\volden~package will be publicly available on the Github repository (\url{https://github.com/aa16oaslak/volden}) upon publication}.

%\bibliography{references}

\bibliography{volden_manuscript_with_track_changes}

\end{document}